\documentclass[journal]{IEEEtran}

\usepackage[utf8]{inputenc}
\usepackage[T1]{fontenc}
\usepackage{hyperref}
\hypersetup{colorlinks=true, linkcolor=blue, filecolor=magenta, urlcolor=cyan}
\usepackage{graphicx}
\usepackage[export]{adjustbox}
\graphicspath{{./images/}}
\usepackage{amsmath}
\usepackage{amsfonts}
\usepackage{amssymb}
\usepackage[version=4]{mhchem}
\usepackage{stmaryrd}
\usepackage{bm}
\usepackage{color,soul}
\usepackage{enumerate}
\usepackage{subcaption}
\usepackage{cite}
\usepackage{microtype}  
\begin{document}
	
	\title{Experimental Verification of a Time-Domain Load Identification Method for Single-Phase Circuits}
	
	\author{Francisco M. Arrabal-Campos,~\IEEEmembership{Member,~IEEE}, 
		Francisco G. Montoya,~\IEEEmembership{Member,~IEEE},
		Jorge Ventura, 
		Santiago Sánchez-Acevedo,
		Raymundo E. Torres-Olguin,~\IEEEmembership{Member,~IEEE}, 
		and~Francisco de León,~\IEEEmembership{Life Fellow,~IEEE}
		\thanks{This work was supported in part by European Research Infrastructure supporting Smart Grid and Smart Energy Systems Research, Technology Development, Validation and Roll Out. Grant H2020, ERIGrid 2.0, GA 870620.}
		\thanks{F. Arrabal-Campos, J. Ventura and F. G. Montoya are with the Department of Engineering, University of Almeria, La Cañada de San Urbano, 04120 Almeria, Spain (e-mail: fmarrabal@ual.es; pagilm@ual.es).}
		\thanks{S. Sánchez-Acevedo and R. E. Torres-Olguin are with SINTEF, Norway (santiago.sanchez@sintef.no, raymundo.torres-olguin@sintef.no).}
		\thanks{F. de León is with the Department of Electrical and Computer Engineering at New York University, Five Metrotech Center, Brooklyn, NY 11201 USA (fdeleon@nyu.edu).}}
	
	\maketitle
	
	\begin{abstract}
		This paper presents experimental validation of a time-domain load parameter determination method for single-phase circuits. The verification is performed in a state-of-the-art smart grid laboratory equipped with power hardware and real-time emulators. The proposed method enables the identification of circuit parameters using only instantaneous voltage and current measurements at the point of common coupling. The experimental setup includes a range of test cases covering linear and non-sinusoidal single-phase conditions. Voltage and current waveforms are acquired, preprocessed, and used to calculate the relevant circuit parameters. The experimental results demonstrate a high degree of accuracy and robustness, with minimal percentage errors across all test cases. The identified parameters show excellent agreement with the theoretical expectations, confirming the validity and applicability of the proposed method to identify the load of single-phase systems. This validation highlights the potential of the method for improved monitoring, control, and protection of smart grids, paving the way for future extensions to three-phase systems and real-time implementations.
	\end{abstract}
	
	\begin{IEEEkeywords}
		Circuit parameter identification, nonlinear circuits, non-sinusoidal excitation.
	\end{IEEEkeywords}

    \section{Introduction}
    Modern Electric Power Systems (EPS) have evolved significantly from systems with simple unidirectional power flow to highly complex networks with distributed resources. In addition, the development of electronic systems and computing power has driven the development of sophisticated Information Technology (IT), which has become integrated with power systems. The coupling of EPS and IT is known as Cyber-Physical-Power Systems (CPPS) \cite{CPPS1,CPPS2,CPPS3}. CPPS are increasing the amount of computational control, protection, or energy management methods that require accurate knowledge of the system parameters, such as resistance, inductance, and capacitance of the EPS's loads or lines. 
    
    Estimation of electric circuit parameters in AC applications is crucial for effective control, protection, and energy management. Accurate parameter estimation are indeed key for enabling advanced grids services such as voltage control \cite{goldin2021time}, congestion management and frequency regulation. Moreover, accurate estimation of parameters is essential for protection systems in both distribution or transmission levels, as  as load characteristics directly influence short-circuit currents, inrush behavior, and transient responses—making load models critical for reliable fault analysis and protection coordination\cite{protection1}. Accurate estimation of the equivalent grid circuits connected to converter-based  systems is crucial for enhancing stability margins of power electronic control systems \cite{control1}.  In energy management based on integrated energy systems, parameter estimation enables optimal power flows management by capturing load dynamics \cite{Emanagement1}.
    
    	
    
    In this context of increasing complexity of modern power systems, reliable methods for circuit parameter identification become essential. To address this demand, a time-domain load parameter identification method for general linear and non-linear single-phase loads was published by a subset of the authors of this paper in 2022 \cite{montoya2022}. Although the theoretical foundation of this approach was showcased through computer simulations, its practical implementation must be validated under real-world conditions. This validation is particularly important given the challenges in parameter estimation for modern power systems  with non-linear loads and non-sinusoidal excitations. This paper presents a laboratory-based experimental verification of the proposed method as a critical step toward its future real-time implementation in the field.

    The problem  addressed through simulation in \cite{montoya2022}  is experimentally verified in this paper. 
    Based solely on measurements of instantaneous voltage and current at the Point of Common Coupling (PCC), the method is capable of identifying the discrete (and possibly time dependent) $RLC$ parameters of equivalent series or parallel circuits. In other words, the method can identify the resistor $R$, inductor $L$, and elastance $S$ (i.e. the inverse of a capacitor) elements of a series equivalent circuit or the conductance $G$, compliance $\Gamma$ (i.e. the inverse of an inductor), and capacitor $C$ of a parallel equivalent circuit. 
    
    This paper presents two significant contributions. First, it introduces a practical methodology for the experimental calculation of parameters, specifically addressing the issue of signal noise in the original voltage and current measurements. This novel approach avoids problems associated with parameter estimation associated with ill-posedness in circuit parameter estimation. The second contribution is the experimental validation of the proposed methodology. Prior to this work, the method had not been tested on a physical system. Furthermore, the paper demonstrates how the results obtained in \cite{montoya2022}, which are based on differential geometry and geometric algebra, can also be obtained using a system representation of linear differential function similar to the Wronskian matrix.

    
	\section{Theoretical Background}
	\label{sec:theory}
	The rigorous mathematical derivation of the identification method was published in \cite{montoya2022} using Geometric Algebra (GA). While geometric algebra is a powerful tool for vector operations, it remains quite unfamiliar to many engineers. This section provides a simplified derivation of the identification expressions using only differentiation of KVL (Kirchhoff Voltage Law) and KCL (Kirchhoff Current Law) to a set of series or parallel equivalent circuits. The following assumptions are made:
	
	\begin{enumerate}[a)]
		\item Given that modern instrumentation operates digitally, parameter identification of the load is performed over discrete time intervals rather than continuously. Consequently, all non-linearities are modeled as time-dependent discrete functions.
		\item The $RLC$ parameters of the equivalent circuits are assumed to remain constant over a time window longer than the sampling period of the measurements. This ensures that the resulting systems of equations are linearly independent, as discussed in \cite{leon2010}.
	\end{enumerate}
	
	\subsection{Series Equivalent Circuit}
    Let $v$ and $i$ be the instantaneous voltage and current measured at the PCC at a constant sampling rate. Applying KVL to a simple $RL$ circuit yields:
	
	\begin{equation}
		v=Ri+Li'
		\label{eq:RL_circuit_equation}
	\end{equation}
	
	\noindent where $i'$ is the time derivative of the current $i$. Taking the first and second derivatives of \eqref{eq:RL_circuit_equation} yields:
	
	\begin{equation}
		\begin{aligned}
			v' &= Ri' + Li'' \\
			v'' &= Ri'' + Li'''
		\end{aligned}
		\label{eq:RL_circuit_derivatives}
	\end{equation}
	
	The set of equations in \eqref{eq:RL_circuit_derivatives} can be written in matrix form to compute $R$ and $L$ as:
	
	\begin{equation}
		\begin{bmatrix}
			i' & i'' \\
			i'' & i'''
		\end{bmatrix}
		\begin{bmatrix}
			R \\
			L
		\end{bmatrix} =
		\begin{bmatrix}
			v' \\
			v''
		\end{bmatrix}
		\label{eq:RL_equations_matrix}
	\end{equation}
	
	The analytical solution of \eqref{eq:RL_equations_matrix} yields:
	
	\begin{equation}
			R = \frac{v'i''' - i''v''}{i'i''' - (i'')^2} \qquad
			L = \frac{i'v'' - v'i''}{i'i''' - (i'')^2}
			\label{eq:RL_serie_solution}
	\end{equation}
	
	Equation \eqref{eq:RL_serie_solution} is identical to equation (17) in \cite{montoya2022}, where it was rigorously derived using GA. A more detailed model with a capacitor in series can also be obtained by introducing a third derivative. This extension can only be computed under nonsinusoidal voltage supply, which is a common condition in real applications.  
	For mathematical convenience, the elastance $S$ is used instead of the capacitance itself in the KVL expression, resulting in the following formulation:
	
	\begin{equation}
		\begin{aligned}
			v' &= Si + Ri' + Li'' \\
			v'' &= Si' + Ri'' + Li''' \\
			v''' &= Si'' + Ri''' + Li''''
		\end{aligned}
	\end{equation}
	
	\noindent then, we can build the following set of equations:
	
	\begin{equation}
		\begin{bmatrix}
			i & i' & i'' \\
			i' & i'' & i''' \\
			i'' & i''' & i''''
		\end{bmatrix}
		\begin{bmatrix}
			S \\
			R \\
			L
		\end{bmatrix} =
		\begin{bmatrix}
			v' \\
			v'' \\
			v'''
		\end{bmatrix}
		\label{eq:RLC_series_matrix}
	\end{equation}
	
	Equation \eqref{eq:RLC_series_matrix} can be solved analytically yielding equation (25) of \cite{montoya2022} obtained with GA.
	
	\subsection{Parallel Equivalent Circuit}
	A similar process is applied to  the parallel circuit by using KCL. For a more straightforward derivation, it is convenient to use conductance  $G$ and compliance $\Gamma$. The resulting expressions for the parameter identification of equivalent circuits with two and three elements are:
	\begin{equation}
		\begin{bmatrix}
			v' & v'' \\
			v'' & v'''
		\end{bmatrix}
		\begin{bmatrix}
			G \\
			\Gamma
		\end{bmatrix} =
		\begin{bmatrix}
			i' \\
			i''
		\end{bmatrix}
		\label{eq:RL_parallel_matrix}
	\end{equation}
	\begin{equation}
		\begin{bmatrix}
			v & v' & v'' \\
			v' & v'' & v''' \\
			v'' & v''' & v''''
		\end{bmatrix}
		\begin{bmatrix}
			C \\
			G \\
			\Gamma
		\end{bmatrix} =
		\begin{bmatrix}
			i' \\
			i'' \\
			i'''
		\end{bmatrix}
				\label{eq:RLC_parallel_matrix}
	\end{equation}
    The analytical solution of \eqref{eq:RL_parallel_matrix} and \eqref{eq:RLC_parallel_matrix} yields the expressions derived in \cite{montoya2022} using GA.  It is important to note that while three-element models are particularly advantageous when both voltage and current exhibit non-sinusoidal behavior, the methodology offers considerable flexibility in terms of model selection. The choice of circuit representation—whether series, parallel, or a hybrid configuration—is determined by the requirements of the specific application rather than any limitations of the identification method itself. 

    \subsection{Other Models}
    While series and parallel $RL$ configurations represent many practical power systems with acceptable accuracy, the methodology presented in this paper is not limited to these basic forms and and extends to a wide range of circuit topologies. The parameter identification framework can be generalized to accommodate various combinations of resistive, inductive, and capacitive elements in both series and parallel arrangements. 

    Beyond the fundamental configurations presented in previous sections, the method can be applied to more complex structures, such as hybrid parallel-series topologies (e.g., a resistor in parallel with an $RL$ series branch), $RC$ networks, and T or $\Pi$ equivalent circuits commonly used in the modeling of transmission line, transformers, motors, etc. The mathematical framework remains consistent—differentiating the governing equations (KVL or KCL) and establishing a system of linear equations for parameter identification.

    The flexibility of this approach lies in its ability to adapt to different levels of model complexity based on specific application needs. While simpler models offer computational efficiency and are often sufficient for many practical scenarios, more elaborate configurations can be employed when higher fidelity is required. This adaptability is particularly valuable in systems exhibiting nonlinear behavior or operating under non-sinusoidal conditions, where simplified models may not fully capture the system dynamics.

	\subsection{Numerical Calculation of Derivatives}

    We present a numerical differentiation approach using Finite Impulse Response (FIR) filters, chosen for their stability and linear phase properties. By imposing coefficient symmetry \cite{PAQUELET201857} and specific constraints for each differentiation order \cite{TSENG20121317}, we optimize these filters to approximate the ideal differentiator \(H_d(\omega)=j\omega\) at low frequencies while converging to zero as \(\omega \rightarrow \pi\) \cite{robust_differenciator}.

    The numerical derivative at point \(x^*\) is approximated via symmetric differences:
    \begin{equation}
            f'(x^*) \approx \frac{1}{h}\sum_{k=1}^{M} c_k \left[f(x^*+kh)-f(x^*-kh)\right]
    	\label{eq:num_derivative_point}
    \end{equation}   
    where $h$ is the sampling interval, $M=\frac{N-1}{2}$ (with $N$ odd), and $c_k$ are coefficients determined by imposing tangency conditions on the frequency response $2j\sum_{k=1}^Mc_ksin(k\omega)$.
    
    For each order of differentiation, we derive specific constraints through tangency conditions at low frequencies and boundary behavior at $\omega=\pi$. For the ideal first derivative ($H_d(\omega)=j\omega$), second derivative ($H_d(\omega)=-\omega^2$) and third derivative ($H_d(\omega)=-j\omega^3$), these constraints are expressed as:   
    \begin{equation*}
    \begin{aligned}
    &\text{First order:} & 2\sum_{k=1}^{M} c_k k &= 1, & 2j\sum_{k=1}^{M} c_k k\, (-1)^k &= 0 \\
    &\text{Second order:} & 2\sum_{k=1}^{M} c_k k^2 &= 2, & 2\sum_{k=1}^{M} c_k k\, (-1)^k &= 0 \\
    &\text{Third order:} & 2\sum_{k=1}^{M} c_k k^3 &= 6, & 2j\sum_{k=1}^{M} c_k k\, (-1)^k &= 0 
    \end{aligned}
    \end{equation*}     
    These constraints generate systems of linear equations that can be solved analytically, providing robust estimation of discrete derivatives, particularly effective in noisy environments.

	\section{Experimental Setup and Data Acquisition}
    \label{sec:experimental}
	\subsection{Laboratory Infrastructure and Test Setup}
	
\begin{figure}[!t]
		\centering
		\includegraphics[max width=\columnwidth]{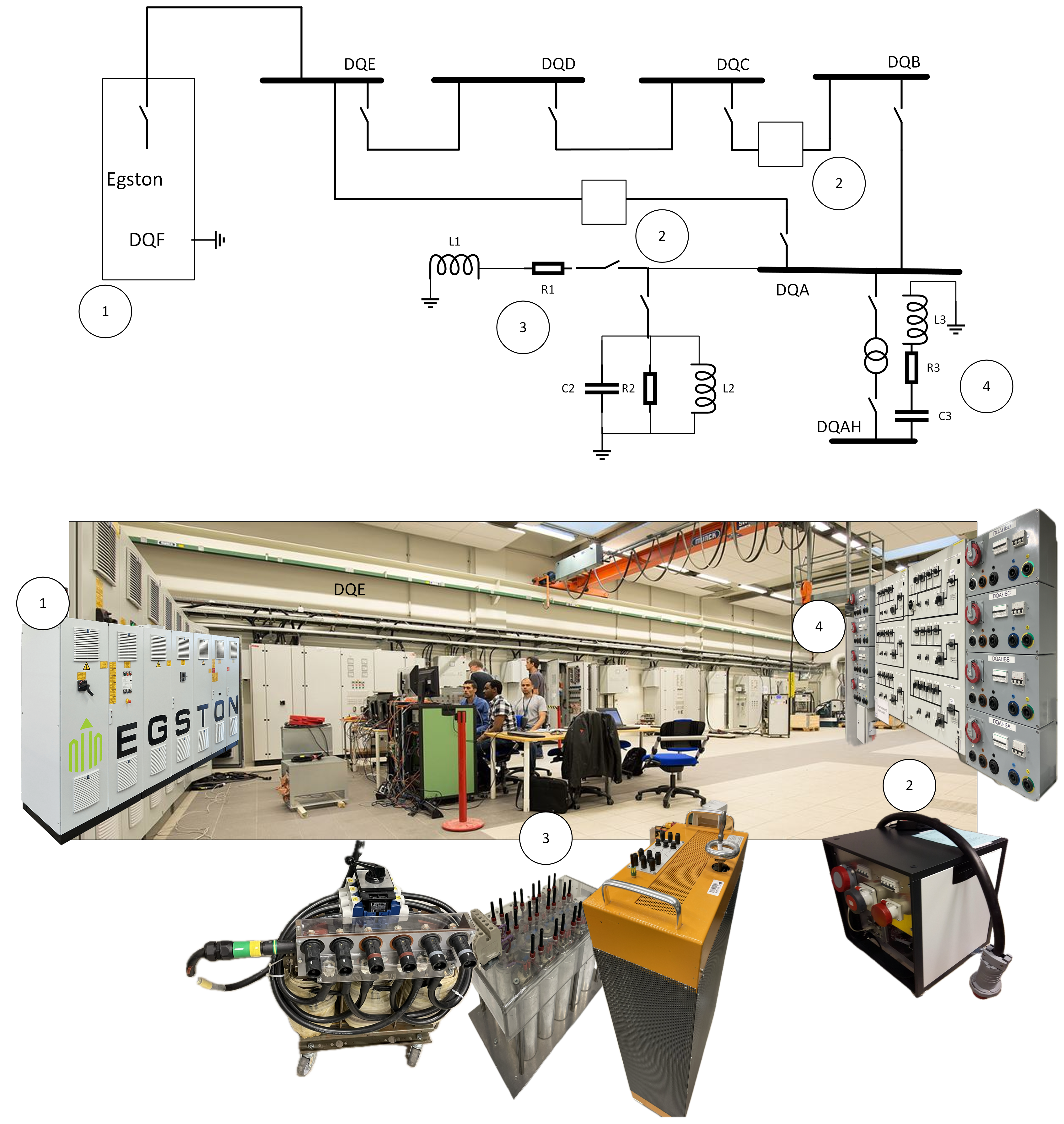}
		\caption{Schematic diagram of SINTEF's laboratory infrastructure showing the main components and their interconnections.}
		\label{fig:lab_schematics}
	\end{figure}

	The experimental validation was conducted at the National Smart Grid Laboratory (SINTEF) in Trondheim, Norway, using a state-of-the-art smart grid testing environment \cite{CPPS1}. As shown in Fig. \ref{fig:lab_schematics}, the laboratory's core infrastructure is built around a 200 kVA power amplifier EGSTON that enables precise voltage waveform generation with 2 kHz bandwidth. This system interfaces with an OPAL-RT real-time simulator platform, providing the capability for sophisticated hardware-in-the-loop testing and real-time data processing.
	
	The measurement system architecture 
    centers around the PCC where high-precision voltage and current sensors connect via isolated channels to Zynq-7000 SoC FPGA PicoZed based acquisition boards. Data acquisition signals are transmitted via a dedicated fiber optic communication network, ensuring high-speed data transfer with minimal interference. Multiple configurable power buses, designated as the DQXX series, provide flexible connection points for various load configurations throughout the laboratory space.
	
	The entire system is monitored and controlled through an integrated MATLAB/Simulink interface coupled with OPAL-RT visualization tools, enabling real-time observation and data collection during experimental runs. This setup ensures precise synchronization between power generation, measurement acquisition, and data processing subsystems, while maintaining signal integrity throughout all testing phases.
	
	\subsection{Test Cases}
	A comprehensive experimental validation test was designed and executed to assess the effectivity of the proposed parameter identification method across a range of circuit configurations and operating conditions. While the overall test plan included a wide range of passive elements and different components, this study focuses on five fundamental test cases that demonstrate the method's capabilities:
	\begin{enumerate}
		\item \textit{Series $RL$ Circuit}: Resistance values ranging from 1.54~$\Omega$ to 19.2~$\Omega$, with fixed inductance of 5~mH. Measurements were taken under both sinusoidal and non-sinusoidal voltage conditions.
		\item \textit{Parallel $RL$ Circuit}: Resistance of 0.77~$\Omega$ in parallel with a series $RL$ branch ($R=1\,\Omega$, $L=5$~mH). Tests performed at different voltage amplitudes and frequencies.
		\item \textit{Parallel $RC$ Circuit}: Fixed resistance of 5.5~$\Omega$ in parallel with a capacitor of 250~$\mu$F. Waveform measurements captured under various operating conditions.
		\item \textit{Distribution Line}: Series $RL$ representation with variable resistance and inductance steps to emulate line parameter changes.
	\end{enumerate}

	All measurements were conducted using the EGSTON power source to generate voltage waveform, and high-precision sensors for data acquisition.
	
 	\section{Measurement Procedure}
 	\label{sec:measurement}
 	\subsection{Data Acquisition}

    Data acquisition was carried out using OPAL-RT with Xilinx Virtex-7 FPGA, which features an embedded PowerPC processor core, providing robust computational capabilities for real-time processing. The measurement system uses the LEM LV 25-1000 voltage transducer, capable of precise voltage measurement up to ±1500 V, while the LEM LF 210-S current transducer enabled accurate bi-directional current sensing up to 200 A. These components ensured high-fidelity signal acquisition in the range of test conditions. Signal synchronization between voltage and current channels was maintained through the OPAL-RT platform's hardware synchronization capabilities, ensuring precise temporal alignment of all measurements.
 	
 	\subsection{Signal Processing}
\begin{figure}[!t]
 		\centering
 		\includegraphics[max width=\columnwidth]{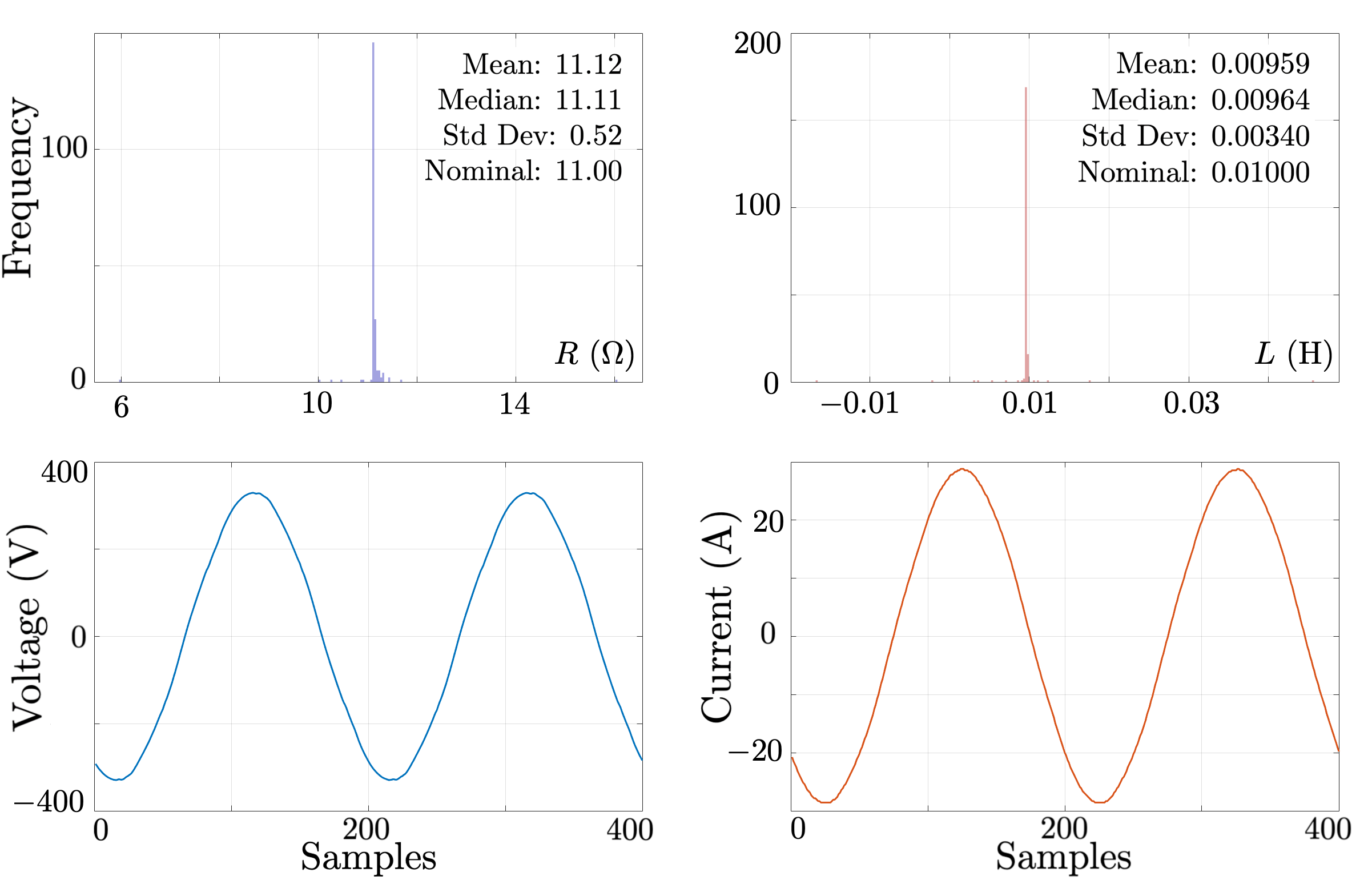}
 		\caption{Filter performance analysis: Resistance and inductance estimation histograms (top), raw voltage and current waveforms (bottom).}
 		\label{fig:filter_analysis}
 	\end{figure}

 	Raw data underwent a comprehensive conditioning pipeline to maximize measurement accuracy. The preprocessing stage began with DC offset removal by subtracting the mean value from both voltage and current signals. A systematic analysis was conducted to determine optimal filtering parameters, evaluating multiple filter types across various configurations. An example is shown in Fig. \ref{fig:filter_analysis} for a specific case. Note that the statistics box in each histogram provides mean, median, standard deviation, and nominal values.
 	
 	The filter analysis encompassed Butterworth, Chebyshev Type I and II, and elliptic filters, evaluated across filter orders (2, 4, 6, 8, and 10) and cutoff frequencies (100 Hz to 2 kHz in 200 Hz steps). The final implementation adopted a Butterworth filter configuration, selected for its maximally flat frequency response and minimal phase distortion characteristics.
 	
 	\subsection{Parameter Calculation Implementation}
 	The parameter identification algorithm processed the filtered signals through the GA framework described in Section \ref{sec:theory}. The implementation computed signal derivatives up to third order using discrete differentiation with appropriate smoothing. A moving average window of 500 samples was applied to the computed parameters to reduce the impact of residual noise while maintaining the ability to track parameter variations.
 	
 	As demonstrated in Fig. \ref{fig:filter_analysis}, the statistical distribution of identified parameters showed excellent agreement with nominal values. For instance, the $RL$ series configuration with nominal values of $R = 11\,\Omega$ and $L = 10$~mH, the median values closely matched specifications with minimal dispersion. These preliminary findings support the validity of the proposed approach, paving the way for a comprehensive analysis of all studied cases presented in Section \ref{sec:results}.
	
	\section{Results and Discussion}
	\label{sec:results}

	\subsection{Basic Circuit Configurations}
	
	\subsubsection{Series and Parallel RL Analysis}
    	\begin{figure}[!t]
		\centering
		\begin{subfigure}{0.48\columnwidth}
			\centering
            \includegraphics[width=\linewidth]{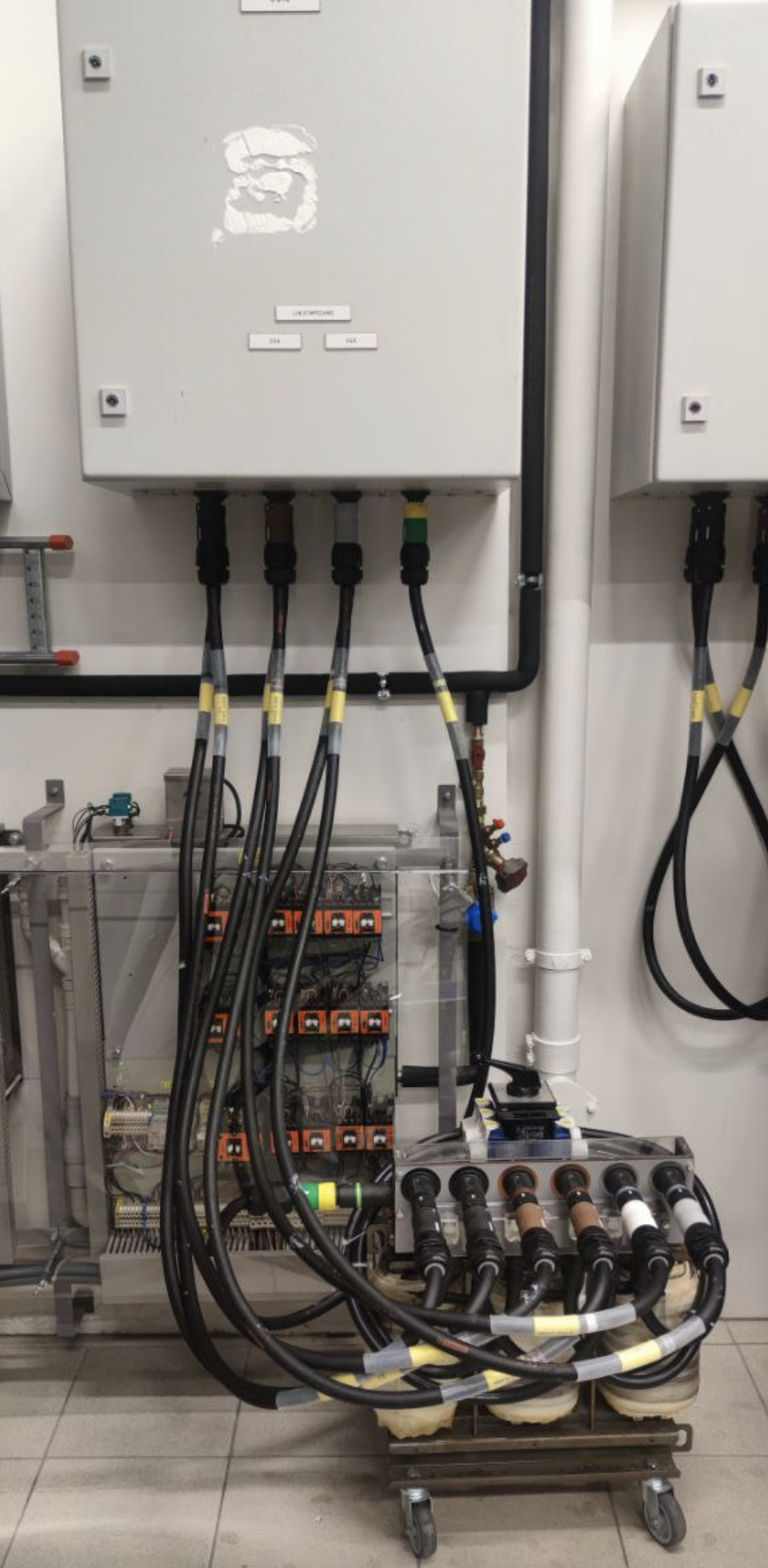}
			\label{fig:series_setup}
		\end{subfigure}
		\hfill
		\begin{subfigure}{0.48\columnwidth}
            \centering
			\includegraphics[width=\linewidth]{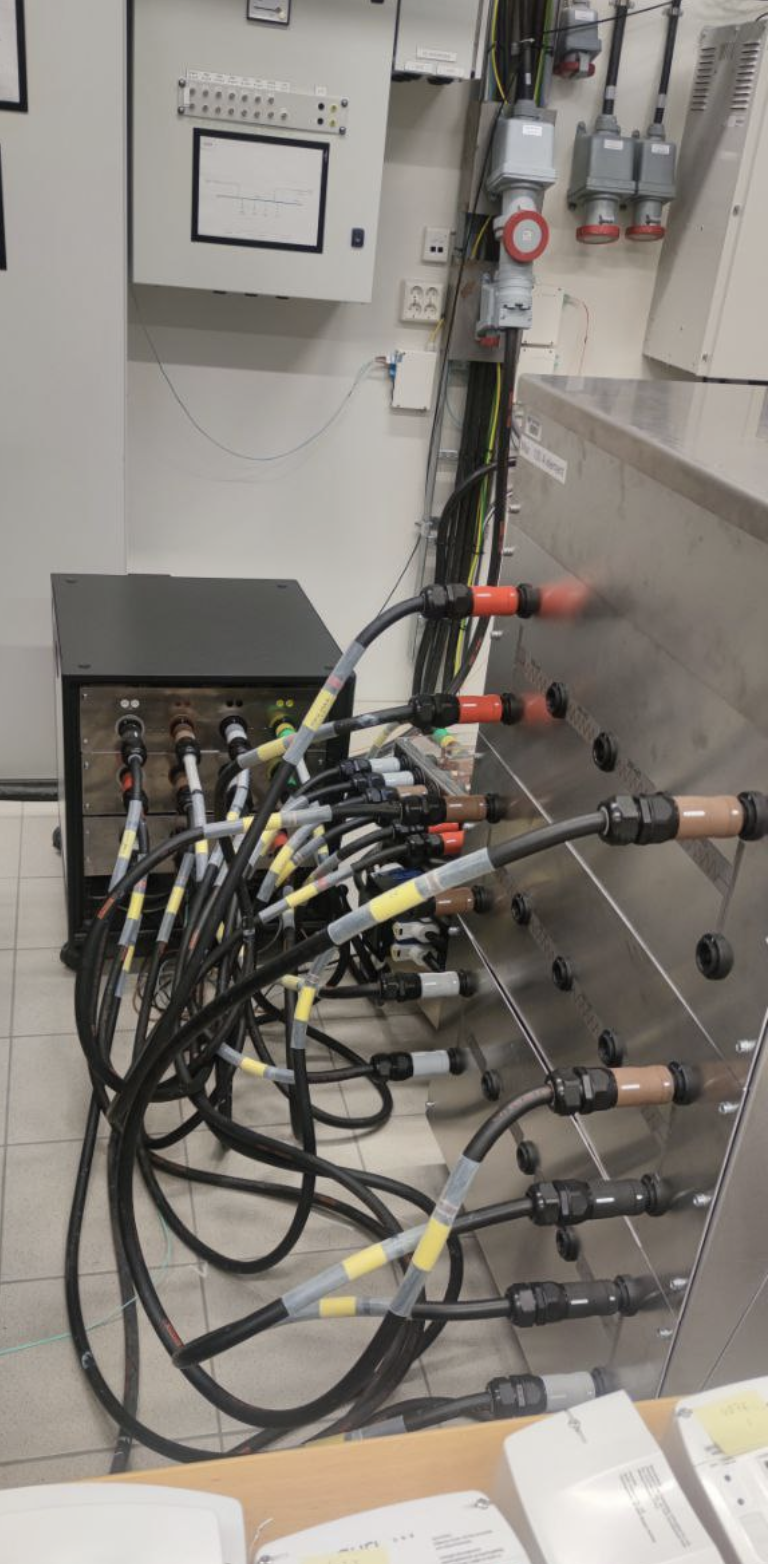}
			\label{fig:parallel_setup}
		\end{subfigure}
		\caption{Laboratory implementation of series and parallel $RL$ circuits at SINTEF facilities.}
		\label{fig:RL_setups}
	\end{figure}
    \begin{figure}[!th]
    	\centering
    	\includegraphics[width=\linewidth]{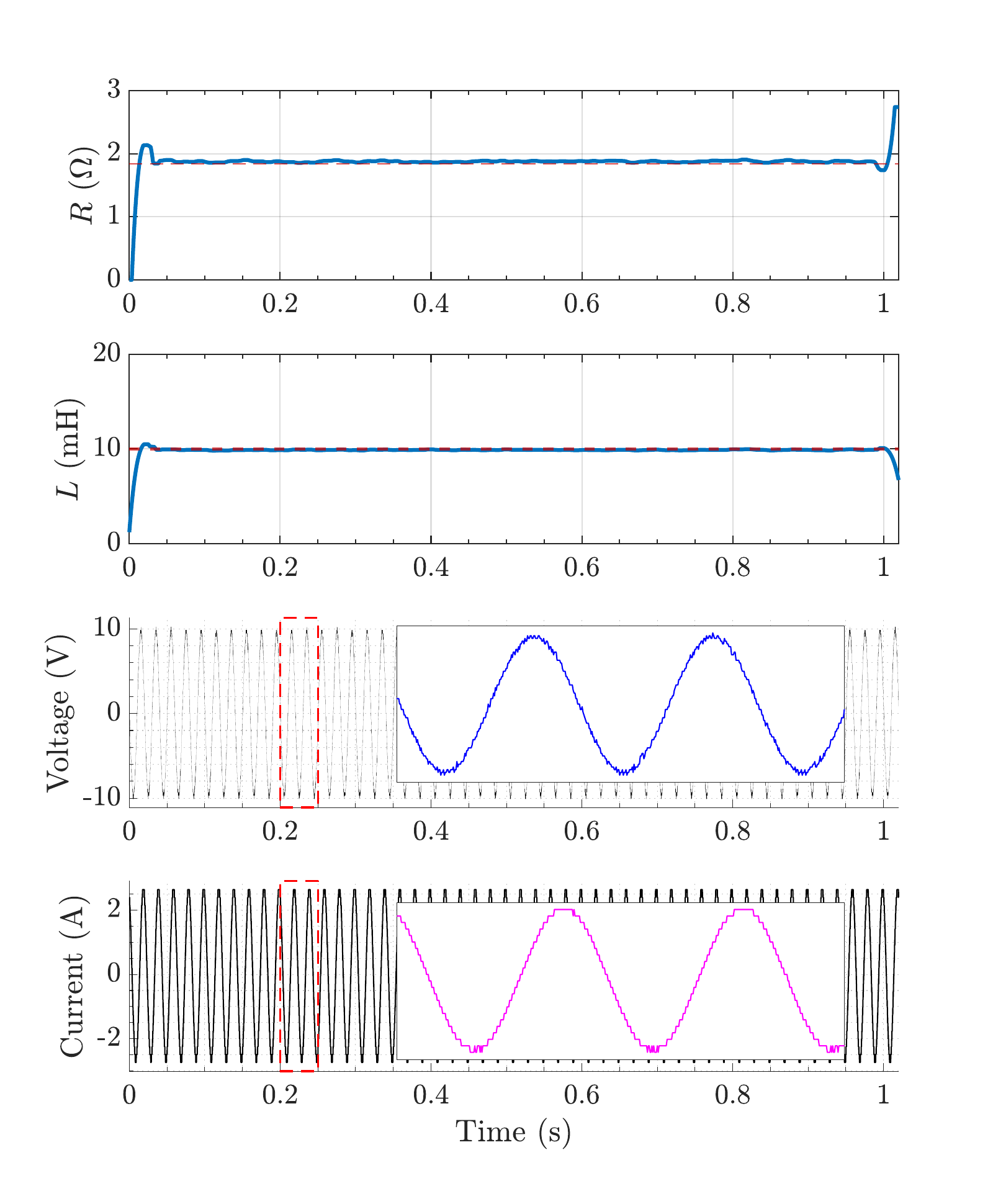}
    	\caption{Identified $R$, $L$, voltage, and current over time in an $RL$ series test (distorted waveforms).}
    	\label{fig:RLfixed}
    \end{figure}
	The experimental validation began with series and parallel $RL$ configurations, which provided a fundamental test of the method's capabilities. Fig. \ref{fig:RL_setups} shows the laboratory implementation of both configurations using the SINTEF infrastructure.
	
	For the series configuration, two types of experiments were conducted. First, a fixed parameter test was performed using a series $RL$ circuit with nominal values of $R = 1.54$~$\Omega$ and $L = 10$~mH (two 5~mH inductors in series with internal resistance of 0.148~$\Omega$ each), where the voltage and current waveforms were monitored over a one-second interval, as shown in Fig. \ref{fig:RLfixed}. The measurement conditions were challenging, with very low voltage levels resulting in non-sinusoidal waveforms for both voltage and current (as evidenced in the detail views). The parameter identification remained stable throughout the experiment, with resistance and inductance estimations matching their nominal values even under these adverse signal quality conditions, where measurements were near the sensors' detection limits. The computed resistance is 1.87~$\Omega$ which is very close (1.8\% difference) to $1.54+0.148\times 2=1.836~\Omega$.
	
	\begin{figure}[]
		\centering
		\includegraphics[width=\linewidth]{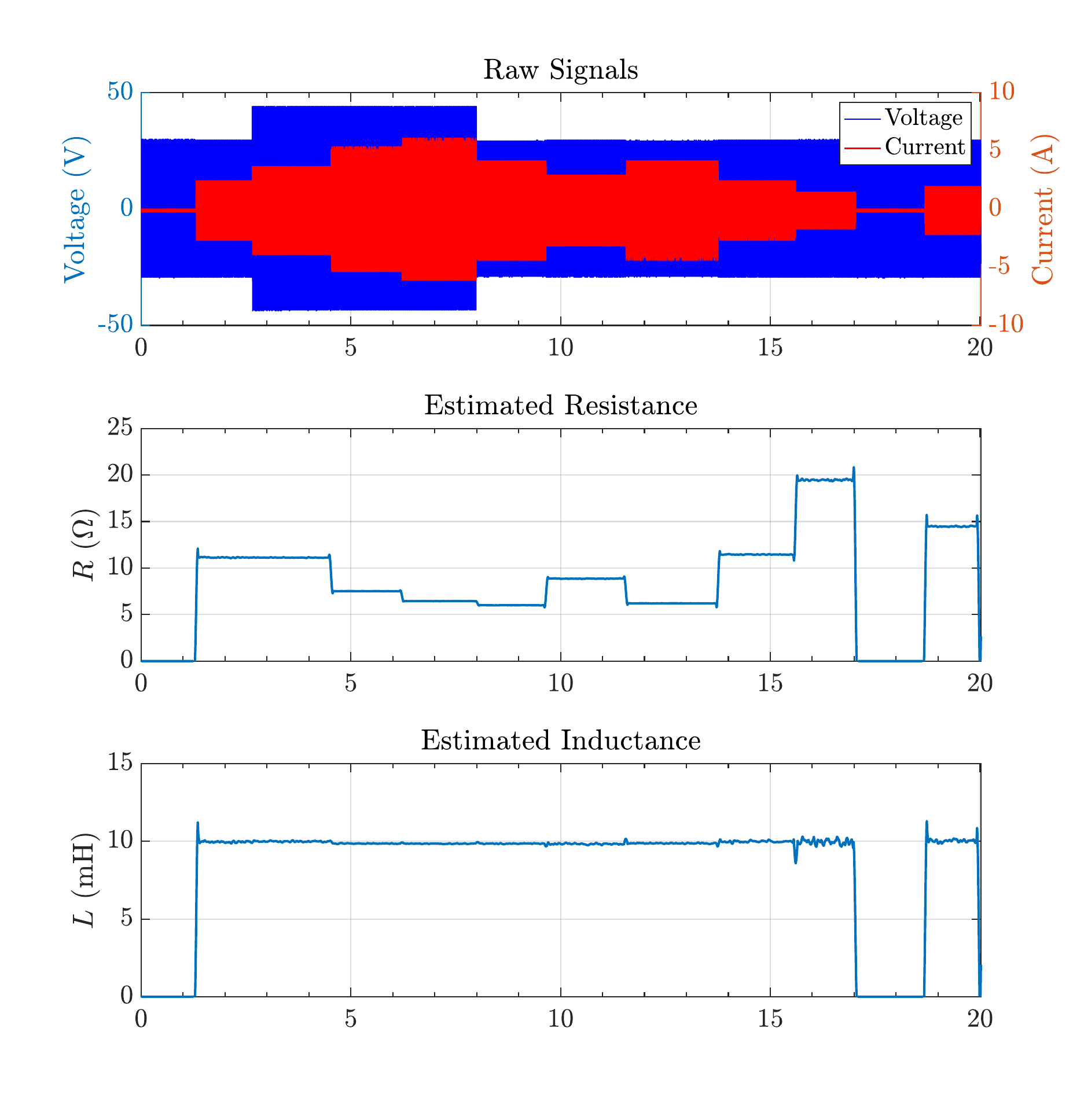}
		\caption{Time evolution of identified parameters and measured signals in the $RL$ series experiment. From top to bottom: Resistance, inductance, applied voltage, and measured current.}
		\label{fig:RLserie1}
	\end{figure}
	
	\begin{figure}[]
		\centering
		\includegraphics[width=\linewidth]{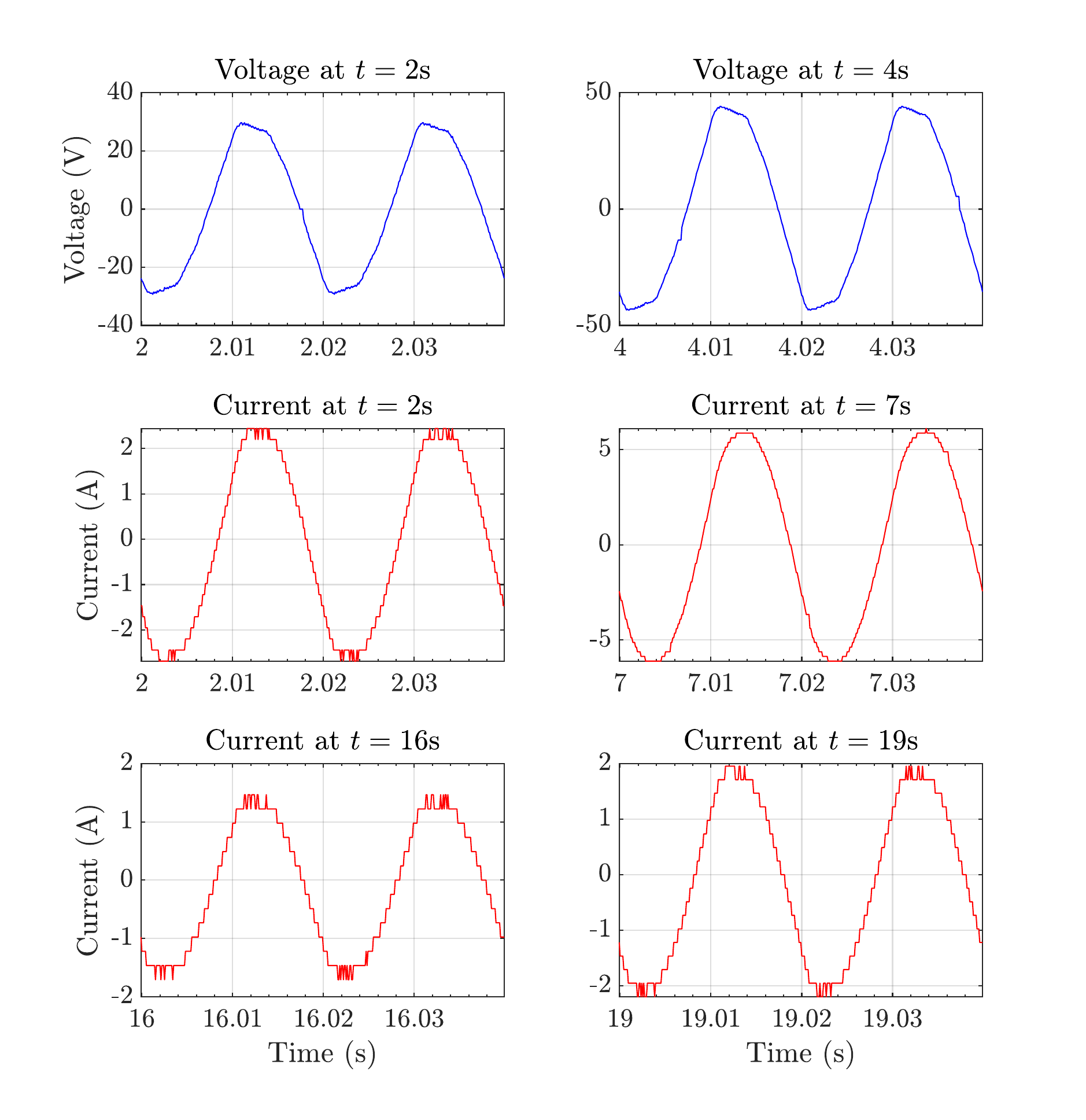}
		\caption{$RL$ series voltage and current waveform details at different time windows.}
		\label{fig:RLserie2}
	\end{figure}
	
	Additionally, variable-parameter tests were conducted with fixed inductance of nominal 10~mH and varying resistance values ranging from 5.5~$\Omega$ to 19.2~$\Omega$. Figs. \ref{fig:RLserie1} and \ref{fig:RLserie2} present an analysis of one experimental run where both voltage amplitude and resistance values were systematically varied. The method identified the inductance value with good results across all test conditions. The resistance identification tracked step changes in resistance values, as evidenced by the clear transitions visible in Fig. \ref{fig:RLserie1} top. The algorithm handled signal distortion well, maintaining parameter identification even under challenging measurement conditions. As shown in Fig. \ref{fig:RLserie2}, this performance was evident in cases where the measured current was so low that ADC quantization effects produced step-like waveforms, while simultaneously handling non-sinusoidal voltage waveforms. This dual capability to process both distorted current and voltage signals demonstrates the method's applicability in real power system scenarios, where ideal waveforms are rarely encountered.
	
\begin{figure}[]
		\centering
		\includegraphics[width=\linewidth]{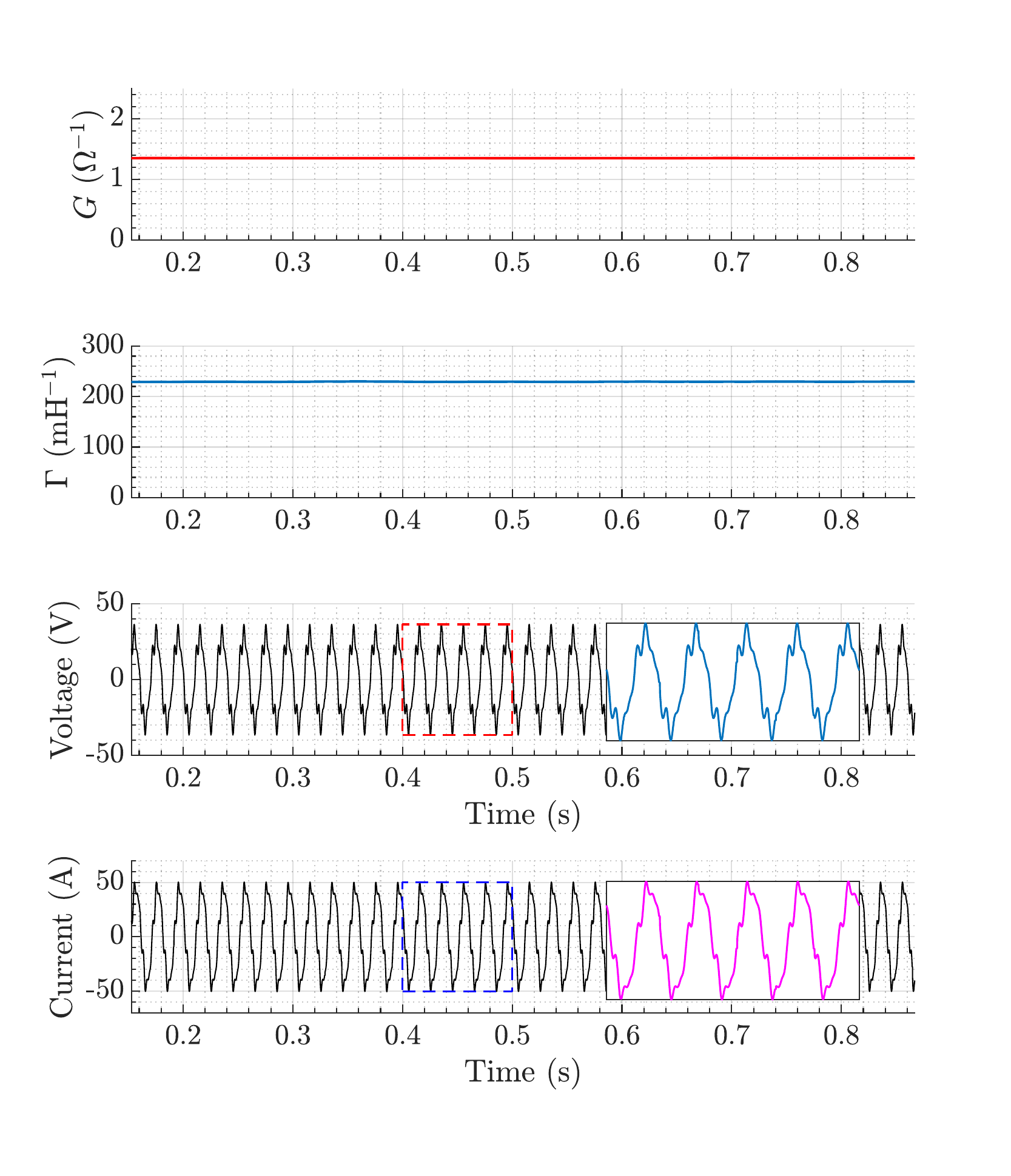}
		\caption{Time evolution of identified parameters and measured signals in the $RL$ parallel experiment. From top to bottom: Conductance, compliance,  voltage, and current.}
		\label{fig:RLparalelo}
	\end{figure}

	For the parallel configuration, a fixed resistance of $0.77\,\Omega$ was connected in parallel with an inductor of $5$~mH. The presence of the inductor's internal resistance (approximately $0.140\,\Omega$) made it challenging to directly obtain the theoretical values of conductance ($G=1/0.77=1.30$~S) and compliance ($\Gamma=1/0.005=200$~H$^{-1}$). However, as shown in Fig.~\ref{fig:RLparalelo}, the method identified constant equivalent parameters ($G\approx 1.35$~S and $\Gamma\approx 229$~H$^{-1}$) that represent the circuit's energy behavior (3.7\% error). These equivalent parameters remain stable throughout the measurement period, indicating the method's ability in handling non-ideal circuit elements. The identified parameters, despite distorted input voltage and current profiles (bottom plots in Fig.~\ref{fig:RLparalelo}), provide an energetic equivalent model of the actual circuit configuration.
	
\begin{figure}[]
		\centering
		\includegraphics[width=\linewidth]{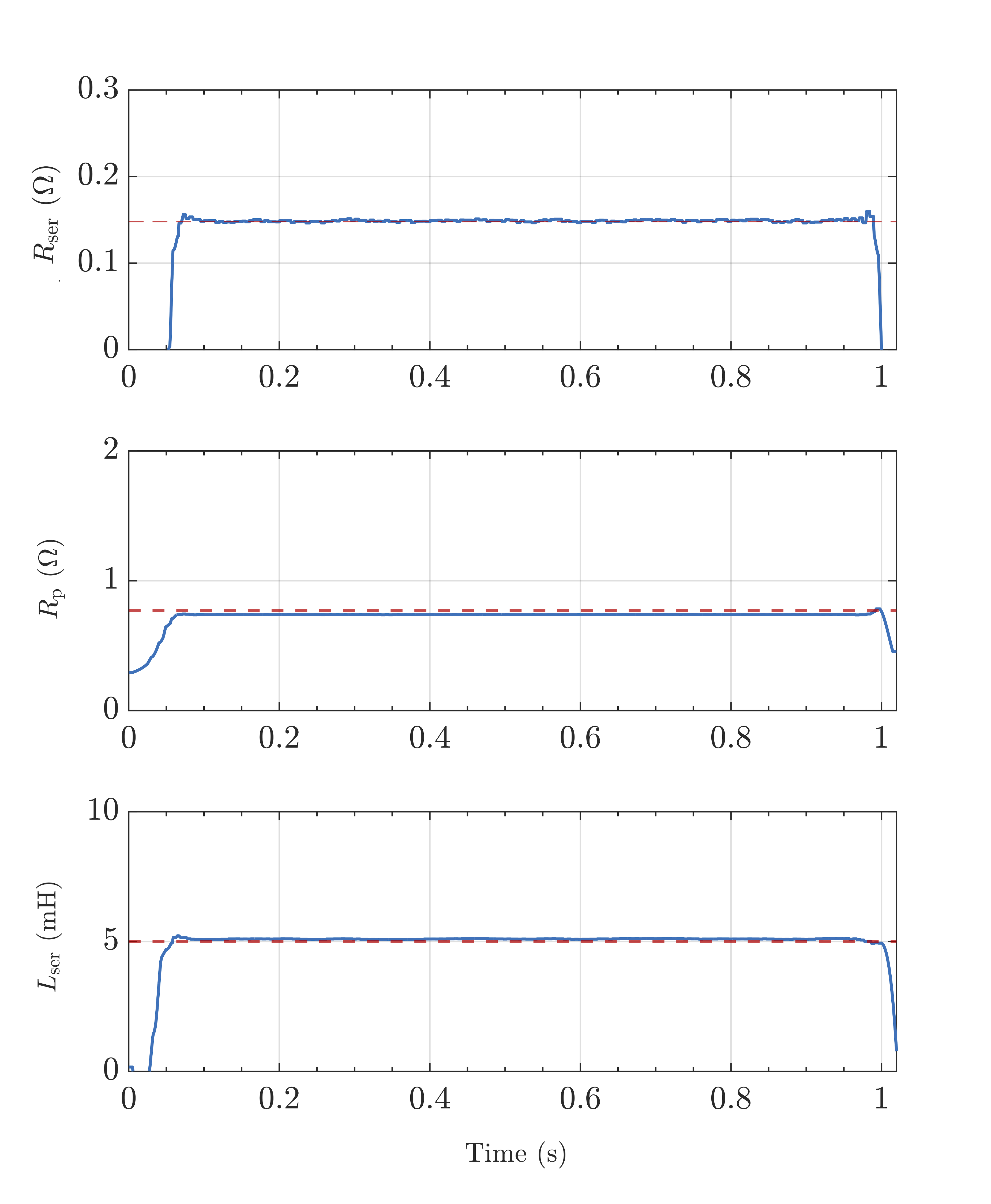}
		\caption{Parameter computation for the $RL$ experiment with advanced identification of the parallel resistor $R_p$, internal resistance $R_{\text{ser}}$ and inductance $L_{\text{ser}}$ of a real inductor.}
		\label{fig:R_RL_parallel}
	\end{figure}

	To improve the parameter identification and better represent the physical reality of the circuit components, a more sophisticated model was subsequently developed. Instead of treating the inductor as ideal, the method was extended to account for its internal resistance by modeling the circuit as a parallel combination of a resistor and an $RL$ series branch. This approach required the computation of third-order derivatives, as shown in the governing equations.	
	\begin{equation*}
		G_p =\frac{a_1}{a_2} \quad
		x =\frac{a_3}{a_4}  \quad
		R_{\text{ser}}=\frac{1}{x-G_p} \quad
		L_{\text{ser}} =-R_{\text{ser}} \frac{a_2}{a_4}
	\end{equation*}
	\begin{equation*}
		\begin{aligned}
			a_1&=i^{\prime \prime}\left(v i^{\prime \prime}-i^{\prime} v^{\prime}\right)-i^{\prime \prime \prime}\left(v i^{\prime}-i v^{\prime}\right)+v^{\prime \prime}\left(i^{\prime 2}-i i^{\prime \prime}\right)\\
			a_2&=i^{\prime \prime}\left(v v^{\prime \prime}-v^{\prime 2}\right)-v^{\prime \prime \prime}\left(v i^{\prime}-i v^{\prime}\right)+v^{\prime \prime}\left(v^{\prime} i^{\prime}-i v^{\prime \prime}\right)\\
			a_3&=i^{\prime \prime}\left(v^{\prime} i^{\prime \prime}-i^{\prime} v^{\prime \prime}\right)-i^{\prime \prime \prime}\left(v^{\prime} i^{\prime}-i v^{\prime \prime}\right)+v^{\prime \prime \prime}\left(i^{\prime 2}-i i^{\prime \prime}\right)\\
			a_4&=i^{\prime \prime \prime}\left(v v^{\prime \prime}-v^{\prime 2}\right)-v^{\prime \prime \prime}\left(v i^{\prime \prime}-i^{\prime} v^{\prime}\right)+v^{\prime \prime}\left(v^{\prime} i^{\prime \prime}-i^{\prime} v^{\prime \prime}\right)
		\end{aligned}
	\end{equation*}	
	This enhanced model improved the accuracy in parameter identification. As shown in Fig.~\ref{fig:R_RL_parallel}, the method estimated the series resistance ($R_{\text{ser}} = 0.148\,\Omega$), parallel conductance ($G_p = 1.354$~S), and series inductance ($L_{\text{ser}} = 5.098$~mH), which align with the nominal values of $0.148\,\Omega$, $1.299$~S, and $5$~mH, respectively. The algorithm shows good stability in parameter identification, with convergence across different operating conditions. The agreement between estimated and nominal values, with errors below 4.2\% for all parameters, indicates the method's capability to model non-ideal elements.
	
	\subsubsection{Parallel RC Configuration}
\begin{figure}[]
		\centering
		\includegraphics[width=0.9\columnwidth]{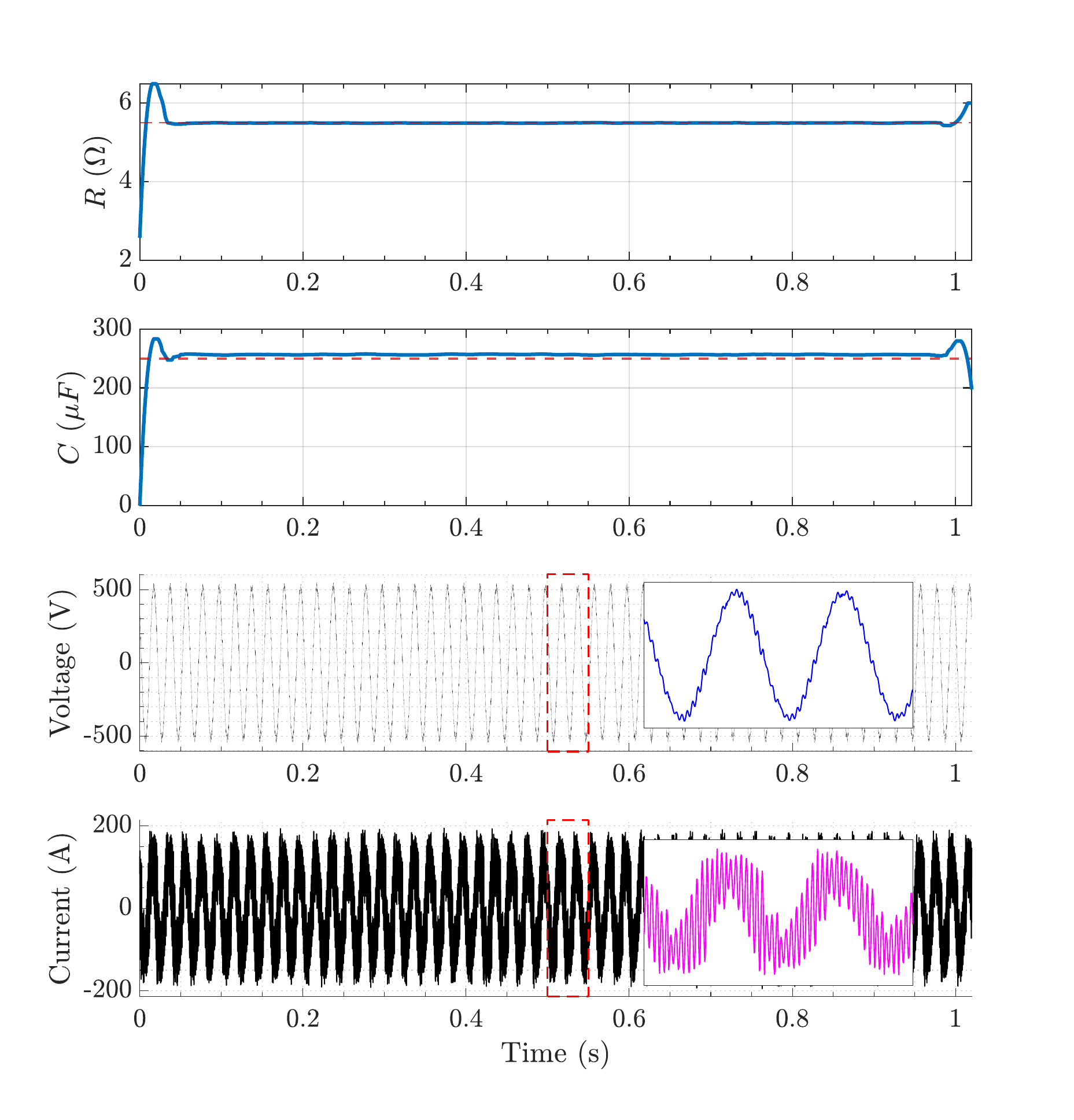}
		\caption{Time evolution of identified parameters and measured signals in the $RC$ parallel experiment. From top to bottom: Resistor, capacitor,  voltage, and current.}
		\label{fig:RC_analysis}
	\end{figure}

    The parallel $RC$ configuration extends the theoretical framework presented in \cite{montoya2022} by deriving new expressions for conductance and capacitance:
    \begin{equation}
    G = \frac{i v^{\prime \prime}-v^{\prime} i^{\prime}}{v v^{\prime \prime}-v^{\prime 2}} \quad 
    C = \frac{v i^{\prime}-i v^{\prime}}{v v^{\prime \prime}-v^{\prime 2}}
    \end{equation}    
    This mathematical formulation, when applied to a parallel combination of a 5.5~$\Omega$ resistor (theoretical conductance of 0.1818~S) and a 250~$\mu$F capacitor, with similar challenging operating conditions. As shown in Fig. \ref{fig:RC_analysis}, the circuit exhibited significant resonance-like behavior, resulting in severe distortion in both voltage and current waveforms. Nevertheless, the method consistently identified a conductance of 0.182~S and a capacitance of 253~$\mu$F, representing errors of less than 0.1\% and 1.2\% respectively from their nominal values. These results remained stable throughout the measurement period, even under highly distorted conditions, validating both the theoretical derivation and the robustness of the identification method.
	It should be noted that during these experiments, temperature variations were not explicitly measured, as tests were conducted at controlled room temperature for relatively short durations. While thermal effects on resistance were not a focus of this study, the method's ability to track parameter changes in real-time makes it potentially valuable for identifying temperature-dependent variations in circuit parameters. This capability could enable indirect temperature estimation in applications where direct thermal measurement is challenging, such as in enclosed power system components or live electrical equipment, representing an interesting direction for future research.
	
	\subsection{Advanced Circuit Configurations}	
	
	\subsubsection{Distribution Line Parameter Estimation}

    \begin{figure}[]
		\centering
		\includegraphics[width=\columnwidth]{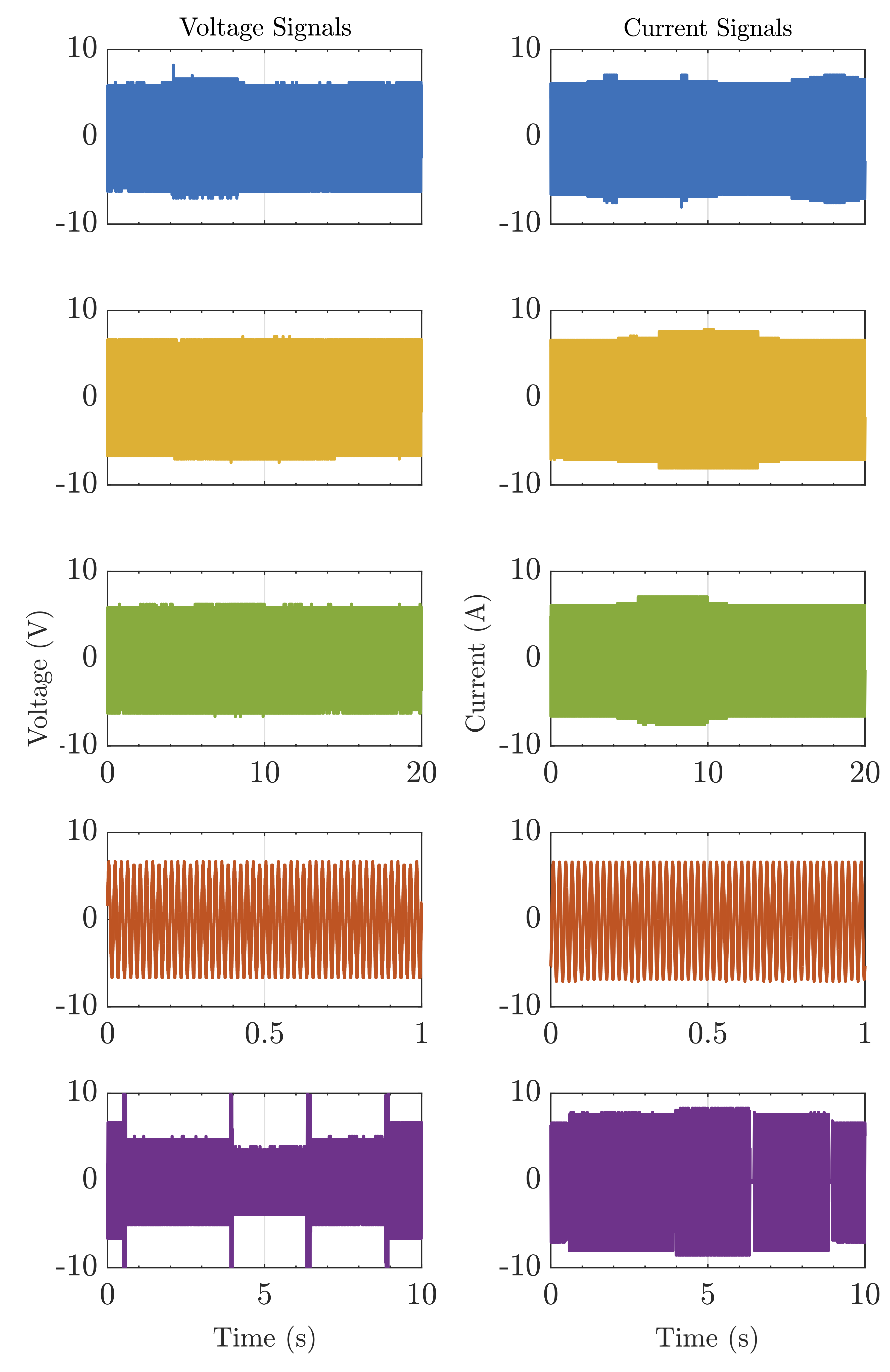}
		    \caption{Voltage and current waveforms for five distribution line test cases, covering resistance steps, frequency variations, steady-state operation, and inductance steps.}
		\label{fig:distribution_line2}
	\end{figure}
    
    \begin{figure}[]
		\centering
		\includegraphics[width=.9\columnwidth]{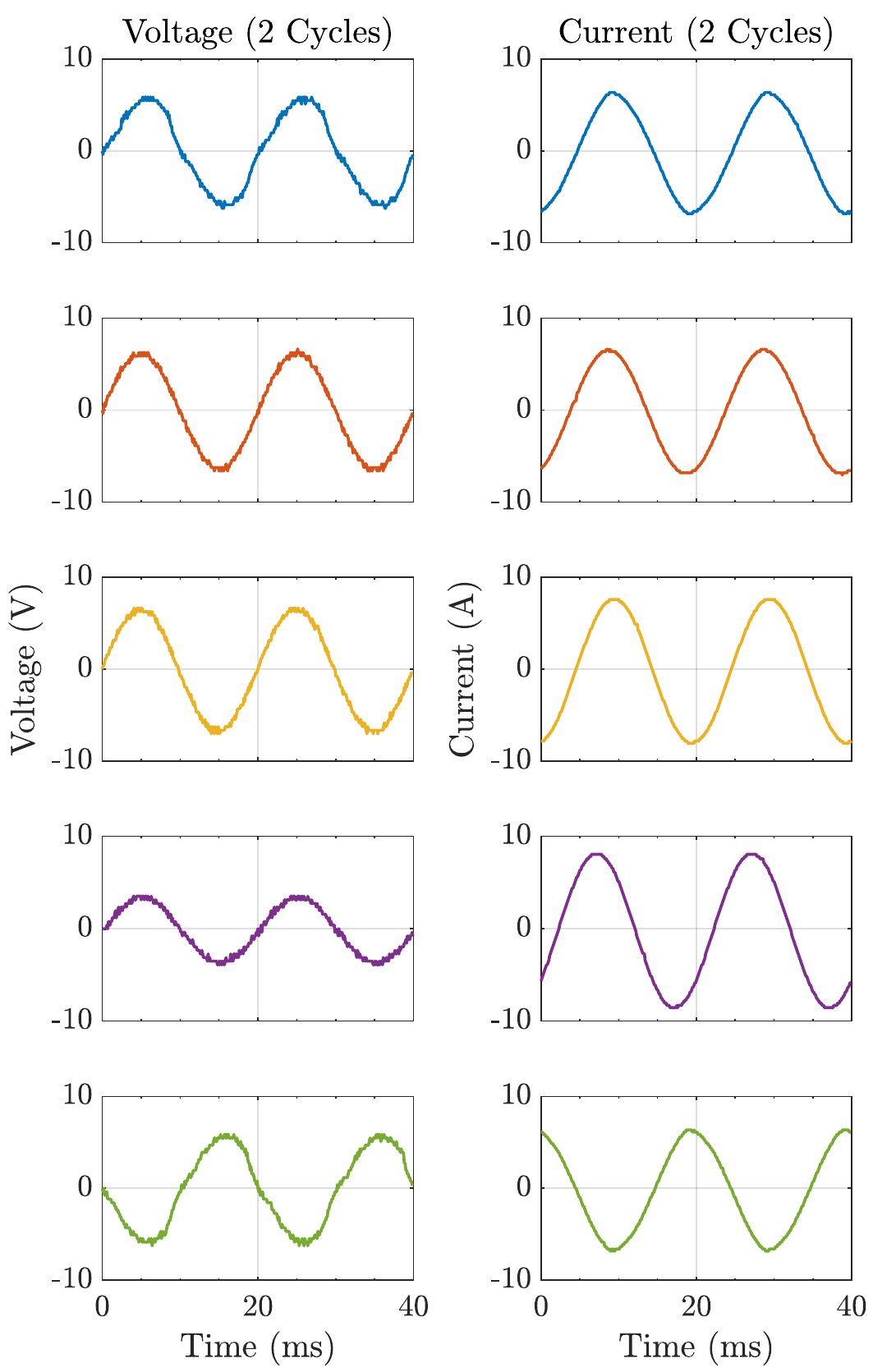}
        \caption{Detailed view of voltage and current waveforms showing distortion characteristics in the distribution line tests.}
		\label{fig:distribution_line3}
	\end{figure}
    
    \begin{figure}[]
		\centering
		\includegraphics[width=.9\columnwidth]{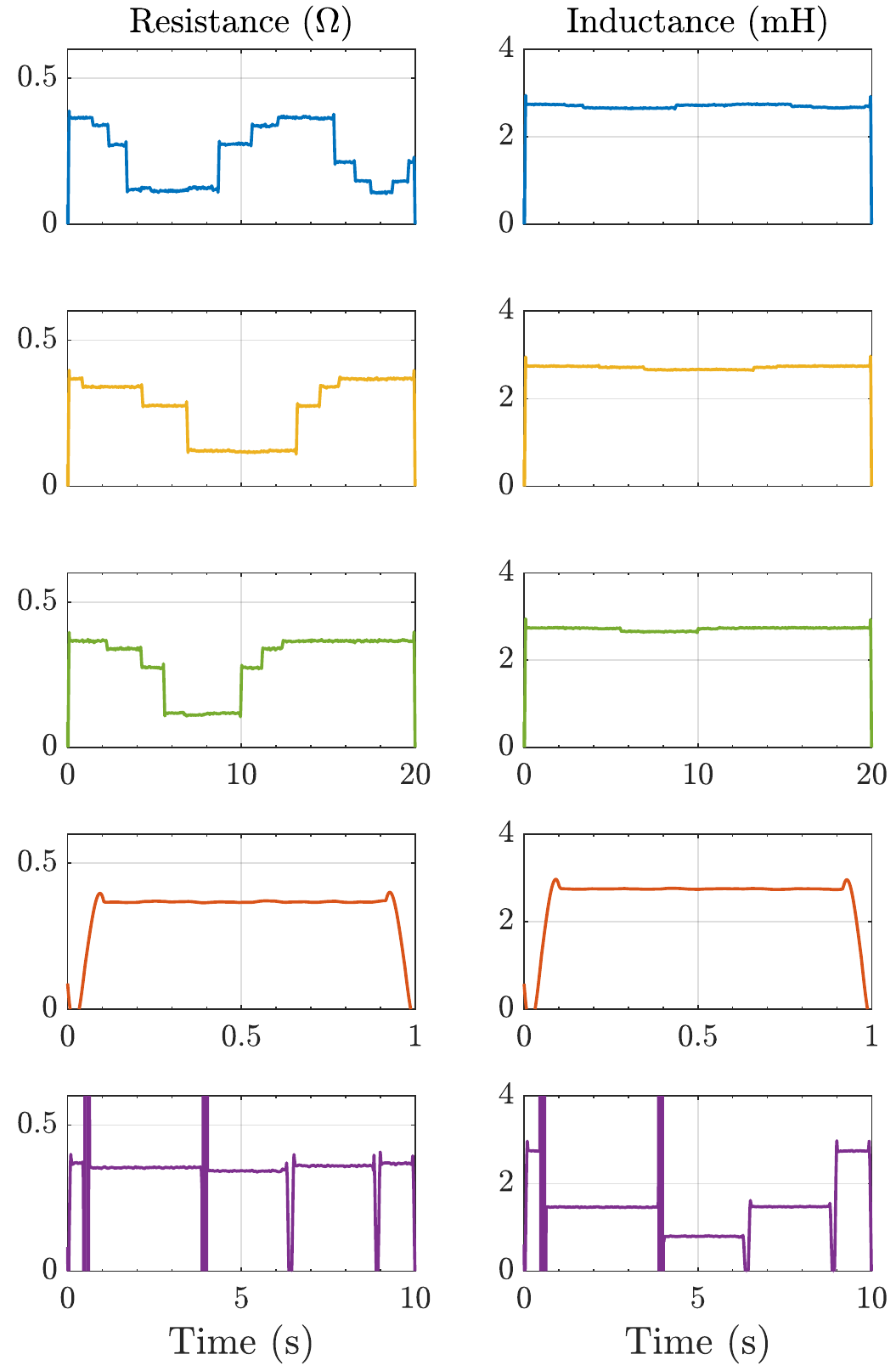}
		\caption{Parameter identification results for distribution lines.
			Left: identified resistances. Right: identified inductances. Rows: (1) resistance steps, (2) combined variations, (3) irregular profile, (4) baseline, (5) inductance steps}
		\label{fig:distribution_line1}
	\end{figure}

    The capability of the proposed method to identify and track time-varying parameters makes it particularly valuable for distribution line monitoring applications. A comprehensive test campaign was conducted using the laboratory infrastructure depicted in Fig. \ref{fig:lab_schematics} to emulate a distribution line under various operating conditions. Five distinct test cases were implemented to evaluate the method's performance across a range of realistic scenarios.
    
    Fig. \ref{fig:distribution_line2} presents the complete voltage (differential measurement between ends) and current waveforms measured at the terminals of the emulated distribution line for all test cases. These measurements capture the entire test sequence, showing the broad range of operating conditions evaluated, including frequency variations. The detailed view in Fig. \ref{fig:distribution_line3} reveals the significant waveform distortion present in the signals, highlighting the challenging conditions under which the algorithm was required to perform. Despite these non-ideal conditions, with the presence of both harmonic content and measurement noise, the parameter identification algorithm performed reasonably well.
    
    The parameter identification results for all five test cases are comprehensively presented in Fig. \ref{fig:distribution_line1}, which displays the estimated resistance (left column) and inductance (right column) values across time for each scenario. The first row corresponds to the test case with resistance steps at fixed frequency, clearly showing the algorithm's ability to track discrete changes in resistance while maintaining stable inductance identification. The second row illustrates combined frequency and resistance variations, demonstrating the method's capability to distinguish between parameter changes and frequency-dependent effects. The third row presents a miscellaneous profile with irregular variations, simulating the complex behavior of distribution lines under varying conditions. The fourth row shows the baseline performance with fixed parameter values, confirming the algorithm's stability during steady-state operation. Finally, the fifth row focuses on inductance steps while maintaining constant resistance, revealing the method's sensitivity to inductive parameter changes.
       
    The experimental validation confirmed the robustness and accuracy of the proposed parameter identification method across all test cases. The method consistently achieved parameter estimation with errors below 4.2\% for both time-invariant and time-variant conditions, maintaining its effectiveness even under highly distorted voltage and current waveforms. A key finding was the method's ability to track rapid parameter changes, as evidenced by the successful identification of resistance variations from 0.77~$\Omega$ to 19.2~$\Omega$ and inductance values around 5~mH with remarkable precision. The algorithm proved particularly stable when processing signals with adequate amplitude, though some sensitivity to measurement noise was observed at very low current levels (below 100~mA, which is far below the sensor capabilities). The identification accuracy remained consistent across different circuit topologies, demonstrating remarkable robustness against waveform distortion, frequency variations, and parameter coupling issues. These comprehensive results highlight the method's versatility and potential for real-time line impedance monitoring in distribution networks, where accurate parameter information is crucial for fault location, state estimation, and adaptive protection schemes. The ability to track parameter variations under diverse operating conditions provides significant advantages over traditional methods that rely on offline calculations or periodic manual measurements, particularly in modern grids with increasing levels of distributed generation and dynamic loading patterns.

\section{Conclusions}
    \label{sec:conclusions}
    This paper presents a comprehensive experimental validation of the time-domain load parameter identification method for  single-phase circuits previously proposed in \cite{montoya2022}. The experimental verification, conducted at the National Smart Grid Laboratory (SINTEF) in Trondheim, Norway, confirms the robustness and practical applicability of the theoretical framework in practical settings under real-world measurement conditions.
    
    The method consistently demonstrates high accuracy in identifying circuit parameters using only instantaneous voltage and current measurements at the point of common coupling, with minimal error margins across all test configurations. Its performance remains stable even under challenging measurement conditions including waveform distortion and signal noise. Notably, the approach successfully handles both linear and non-linear loads across different circuit topologies—series $RL$, parallel $RL$, parallel $RC$, and distribution line configurations—establishing its versatility for diverse power system applications. The implementation of specialized numerical differentiation based on FIR filters significantly enhances the method's resilience to measurement noise, addressing one of the primary challenges in practical parameter identification.
    
    Perhaps most significant is the technique's exceptional dynamic performance, effectively tracking parameter variations over time and rapidly converging after step changes in circuit conditions. This capability, essential for real-time monitoring applications, distinguishes the method from traditional approaches that typically rely on steady-state assumptions.
    
    The successful laboratory validation represents an important step toward field applications, indicating that the methodology could provide valuable capabilities for fault detection, impedance monitoring, load characterization, and adaptive protection schemes. Future research will focus on extending the methodology to three-phase systems, investigating its performance in networks with distributed energy resources, and implementing real-time parameter identification for grid monitoring and control applications.
	
	
	\bibliographystyle{IEEEtran}
	\bibliography{references.bib}

\begin{thebibliography}{10}
\providecommand{\url}[1]{#1}
\csname url@samestyle\endcsname
\providecommand{\newblock}{\relax}
\providecommand{\bibinfo}[2]{#2}
\providecommand{\BIBentrySTDinterwordspacing}{\spaceskip=0pt\relax}
\providecommand{\BIBentryALTinterwordstretchfactor}{4}
\providecommand{\BIBentryALTinterwordspacing}{\spaceskip=\fontdimen2\font plus
\BIBentryALTinterwordstretchfactor\fontdimen3\font minus
  \fontdimen4\font\relax}
\providecommand{\BIBforeignlanguage}[2]{{%
\expandafter\ifx\csname l@#1\endcsname\relax
\typeout{** WARNING: IEEEtran.bst: No hyphenation pattern has been}%
\typeout{** loaded for the language `#1'. Using the pattern for}%
\typeout{** the default language instead.}%
\else
\language=\csname l@#1\endcsname
\fi
#2}}
\providecommand{\BIBdecl}{\relax}
\BIBdecl

\bibitem{CPPS1}
S.~Sanchez-Acevedo and S.~D'Arco, ``Towards a versatile cyber physical power
  system testbed: Design and operation experience,'' in \emph{2021 IEEE PES
  Innovative Smart Grid Technologies Europe}, 2021, pp. 1--6.

\bibitem{CPPS2}
R.~V. Yohanandhan, R.~M. Elavarasan, P.~Manoharan, and L.~Mihet-Popa,
  ``Cyber-physical power system (cpps): A review on modeling, simulation, and
  analysis with cyber security applications,'' \emph{IEEE Access}, vol.~8, pp.
  151\,019--151\,064, 2020.

\bibitem{CPPS3}
S.~Poudel, Z.~Ni, and N.~Malla, ``Real-time cyber physical system testbed for
  power system security and control,'' \emph{International Journal of
  Electrical Power \& Energy Systems}, vol.~90, pp. 124--133, 2017.

\bibitem{goldin2021time}
A.~Goldin, E.~Buechler, R.~Rajagopal, and J.~Rivas-Davila, ``Time and voltage
  domain load models for appliance-level grid edge simulation and control,''
  \emph{Electric Power Systems Research}, vol. 190, p. 106750, 2021.

\bibitem{protection1}
M.~Li, Y.~Luo, J.~He, Y.~Zhang, and A.~P.~S. Meliopoulos, ``Analytical
  estimation of mmc short-circuit currents in the ac in-feed steady-state
  stage,'' \emph{IEEE Transactions on Power Delivery}, vol.~37, no.~1, pp.
  431--441, 2022.

\bibitem{control1}
B.~H. Lin, J.~T. Tsai, and K.~L. Lian, ``A non-invasive method for estimating
  circuit and control parameters of voltage source converters,'' \emph{IEEE
  Transactions on Circuits and Systems I: Regular Papers}, vol.~66, no.~12, pp.
  4911--4921, 2019.

\bibitem{Emanagement1}
B.~Chen, Q.~Guo, G.~Yin, B.~Wang, Z.~Pan, Y.~Chen, W.~Wu, and H.~Sun,
  ``Energy-circuit-based integrated energy management system: Theory,
  implementation, and application,'' \emph{Proceedings of the IEEE}, vol. 110,
  no.~12, pp. 1897--1926, 2022.

\bibitem{montoya2022}
F.~G. Montoya, F.~De~Leon, F.~Arrabal-Campos, and A.~Alcayde, ``Determination
  of instantaneous powers from a novel time-domain parameter identification
  method of non-linear single-phase circuits,'' \emph{IEEE Transactions on
  Power Delivery}, vol.~37, no.~5, pp. 3608--3619, 2022.

\bibitem{leon2010}
F.~De~Leon and J.~Cohen, ``Ac power theory from poynting theorem: Accurate
  identification of instantaneous power components in nonlinear-switched
  circuits,'' \emph{IEEE transactions on power delivery}, vol.~25, no.~4, pp.
  2104--2112, 2010.

\bibitem{PAQUELET201857}
S.~Paquelet and V.~Savaux, ``On the symmetry of fir filter with linear phase,''
  \emph{Digital Signal Processing}, vol.~81, pp. 57--60, 2018.

\bibitem{TSENG20121317}
``Design of linear phase fir filters using fractional derivative constraints,''
  \emph{Signal Processing}, vol.~92, no.~5, pp. 1317--1327, 2012.

\bibitem{robust_differenciator}
P.~Holoborodko, ``Smooth noise robust differentiators,''
  http://www.holoborodko.com/pavel/numerical-methods/numerical-derivative/smooth-low-noise-differentiators/,
  2008.

\end{thebibliography}
	
\end{document}